# Enlarged Surface Meshes and Normalization Conditions for Columns and Rows of Matrices in the COSMO Method

O. Yu. Kupervasser[a] and I. P. Kikot'[b]
a Research Computing Center of Moscow State University, Moscow, Russia
b Semenov Institute of Chemical Physics, Russian Academy of Sciences, Moscow, Russia
e-mail: olegup@yahoo.com; irakikotx@gmail.com

Earlier, normalization conditions for the columns of the PCM (Polarized Continuum Model) were determined and a method of enlarged surface meshes was developed. We developed similar methods for the COSMO (**CO**nductor like **S**creening **MO**del). These methods make it possible to introduce larger surface meshes without loss of accuracy and perform fast calculations of the solvation energy and the Born radii in the SGB (**S**urface **G**eneralized **B**orn) method. In addition, the corrections proposed in this work provide a significant enhancement in the accuracy of numerical calculations.

## 1. INTRODUCTION

It is well known that an adequate treatment of the interaction of charged molecules and surrounding solvent (typically) is very important for modeling various chemical systems, in particular, of biomolecules. There are several different approaches to the modeling of the solvent. In this paper, we will focus on models in which the solvent is considered as a continuous medium with uniform dielectric permittivity and the molecule as a region within this environment with a uniform permittivity, too, which experiences a discontinuity at the boundary. With this approach, the solution of electrostatic problem reduces to determining the surface charge induced by introducing a charged molecule into the dielectric. Below, a brief description of two such models, PCM (**P**olarized **C**ontinuum **M**odel) and COSMO (**CO**nductor like **S**creening **MO**del) [2]. We will consider normalization conditions for columns (rows) of matrices in the COSMO [2]. These methods make it possible to obtain the exact values of the diagonal elements of the COSMO symmetric matrix and to ensure that the condition relating the charge inside the cavity with the surface charge are is met. The conditions examined make it possible to develop an approximate method of enlarged surface meshes for the COSMO, similar to the same method developed for the PCM. This method is expected to perform better than the PCM does, since its matrix elements depend only on the coordinates of the surface meshes and is less sensitive to the errors associated with the enlargement of the surface meshes.

The paper is organized as follows. Section 2 gives a brief description of the normalization conditions for PCM matrices and of the method of enlarged surface meshes, developed in [1]. Section 3 describes the normalization conditions for COSMO matrices and derivation of expressions for the diagonal elements of the symmetric matrix. In Section 4, the method of enlarged meshes as applied to COSMO is formulated.

## 2. PCM WITH THE ENLARGED SURFACE MESHES FOR $\varepsilon=\infty$.

The PCM (polarizable continuum model) [1, 3] is a method for finding exact solutions of an electrostatic problem in which point charges representing the charged atoms of the solute molecule are surrounded by a continuous dielectric medium with a given dielectric constant. In this case, the electric field in the dielectric medium is identical to the electric field that would be created in a vacuum by the inside and the surface charges induced by the solute molecule on the surface separating the dielectric solvent and the molecule. Thus, instead of solving the Poisson equation in three dimensions, it suffices to determine the surface charges.

Consider the exact formulation of the problem of the interaction of point charges inside a dielectric cavity with the ambient dielectric. Suppose there are two simply connected region of space separated by a closed surface $\Xi$. The inner region has a dielectric constant $\varepsilon_{in}=1$, whereas the external, a constant $\varepsilon = \infty$. It should be noted that the PCM can also be successfully $\Xi$. applied in the case of a finite dielectric constant of the ambient medium. However, for consistency with the further description, we present it in the limiting case of $\varepsilon = \infty$. In the cavity contains a system of charges $Q_i$ located at points $\mathbf{R}_i$. The interaction of the charges with the dielectric a surface charge of density $\sigma$ is induced at the interface between regions. The electrostatic potential can be expressed as the sum of the potentials of the point charges inside the cavity and surface charges:

$$\varphi(\mathbf{r}) = \sum_i \frac{Q_i}{\varepsilon_{in}|\mathbf{r} - \mathbf{R}_i|} + \int_\Xi \frac{\sigma(\mathbf{r}')}{|\mathbf{r} - \mathbf{r}'|} dS'. \tag{1}$$

The derivative along the normal to the inner surface is

$$\left(\frac{\partial \varphi_{in}(\mathbf{r})}{\partial n}\right) = \left(-\int_\Xi \frac{\sigma(\mathbf{r}')((\mathbf{r}-\mathbf{r}')\cdot \mathbf{n}(\mathbf{r}))}{|\mathbf{r}-\mathbf{r}'|^3} dS' + 2\pi\sigma\right) - \sum_i \frac{Q_i((\mathbf{r}-\mathbf{R}_i)\cdot \mathbf{n}(\mathbf{r}))}{\varepsilon_{in}|\mathbf{r}-\mathbf{R}_i|^3}. \tag{2}$$

The surface integral is singular at $\mathbf{r} = \mathbf{r}'$; in this case, let us define it as

$$\int_\Xi \frac{\sigma(\mathbf{r}')((\mathbf{r}-\mathbf{r}')\cdot \mathbf{n}(\mathbf{r}))}{|\mathbf{r}-\mathbf{r}'|^3} dS' = \lim_{\delta \to 0} \int_{\Xi/(|\mathbf{r}-\mathbf{r}'|<\delta)} \frac{\sigma(\mathbf{r}')((\mathbf{r}-\mathbf{r}')\cdot \mathbf{n}(\mathbf{r}))}{|\mathbf{r}-\mathbf{r}'|^3} dS'. \tag{3}$$

According to the Gauss–Ostrogradsky theorem,

$$4\pi\sigma = \frac{\partial \varphi_{in}(\mathbf{r})}{\partial n}. \tag{4}$$

Combining Eqs. (2) and (4) yields an integral equation for $\sigma$:

$$4\pi\sigma(\mathbf{r}) = \left(-\int_\Xi \frac{\sigma(\mathbf{r}')((\mathbf{r}-\mathbf{r}')\cdot \mathbf{n}(\mathbf{r}))}{|\mathbf{r}-\mathbf{r}'|^3} dS' + 2\pi\sigma(\mathbf{r})\right) - \sum_j \frac{Q_j((\mathbf{r}-\mathbf{R}_j)\cdot \mathbf{n}(\mathbf{r}))}{\varepsilon_{in}|\mathbf{r}-\mathbf{R}_j|^3}. \tag{5}$$

To numerically solve this integral equation, let us perform discretization — divide the surface into N small meshes, which can be considered point charges. Let $q_i$ be the charge of the *i*-th surface mesh, $q = \{q_i\}_{i=1,N}$ the column vector of charges of surface meshes, $S_i$, the area of the *i*-th mesh, , the vector defining the position of the center of the *i*-th mesh, and vector normal to the *i*-th mesh passing through its center and directed towards the dielectric solvent. Let the number of point charges inside the molecule be $M$ and $Q = \{Q_j\}_{j=1,M}$, column vector of charges of the molecule's atoms. Then, Eq. (5) can be written in the matrix form as

$$4\pi q = ((-Aq + 2\pi q) + 2\pi q) + BQ,$$

For convenience, the first integral here is written as $(-Aq + 2\pi q)$. Then, after canceling similar terms, we obtain

$$Aq = BQ, \tag{6}$$

where $A = \{a_{ij}\}$ is a square matrix of size $N \times N$, $B = \{b_{ij}\}$ is a matrix, which depends on the geometry of surface meshes and the positions of atomic charges of the molecule. The elements of the matrix $B$ have the form

$$b_{ij} = -\frac{\mathbf{n}_i \cdot (\mathbf{r}_i - \mathbf{R}_j)}{|\mathbf{r}_i - \mathbf{R}_j|^3} S_i. \tag{7}$$

For the columns of the matrix $B$, the following identity must hold:

$$\sum_{i=1}^{N} b_{ij} = -4\pi. \tag{8}$$

This identity can be used to correct the errors of discretization, which arise because of division of the surface into meshes:

$$b_{ij} \rightarrow -4\pi \frac{b_{ij}}{\sum_{i=1}^{N} b_{ij}}.$$

Using this correction in the program DISOLV [4–6] gave a significant enhancement in accuracy. The elements of the matrix $A$ are given by [1, 4]

$$a_{ij} = \begin{cases} \dfrac{\mathbf{n}_i (\mathbf{r}_i - \mathbf{r}_j)}{|\mathbf{r}_i - \mathbf{r}_j|^3} S_i, & i \neq j, \\ a_{jj} = 4\pi - \sum_{i \neq j} a_{ij}, & i = j. \end{cases} \tag{9}$$

In the geometric sense, $2\pi - \sum_{i \neq j} a_{ij}$, – is the correction for the curvature of the surface ($a_{jj} = 2\pi$ a small flat). Expression (9) yields the normalization condition for the columns of the matrix $A$:

$$\sum_{i} a_{ij} = 4\pi.$$

(10)

From equation (6), by summing all the rows, and then using the identities (8) and (10), we obtain an identity relating the total surface charge with total charge inside the cavity:

$$\sum_{j=1}^{N} q_j = -\sum_{i=1}^{M} Q_i. \tag{11}$$

To speed up PCM calculations, it is advantageous to use the Totrov–Abagyan method with enlarged surface meshes [1]. It is described by the same initial integral equation (6)–(10) as the ordinary PCM, but for significantly larger meshes. Let us introduce these enlarged\ surface meshes as follows. Consider all the surface atoms. Combine all the small surface meshes into groups consisting of surface meshes closest to a given surface atom. Let the surface charge density be the same at all points for each enlarged surface mesh. Then, the PCM equation for enlarged surface meshes in the matrix form becomes:

$$Rq^{big} = EQ, \qquad (12)$$

where $q^{big}$ is the column of charges of enlarged surface meshes, $E = \{e_{ij}\}$ is the matrix of size $N_{\{surf\}} \times M$, $R = \{R_{ij}\}$ is the square matrix of size $N_{\{surf\}} \times N_{\{surf\}}$, and $N_{\{surf\}}$ is the number of enlarged surface meshes. The matrix elements for the enlarged meshes read as

$$\begin{cases} R_{jk} = \dfrac{\left(\sum_{l_j}\sum_{m_k} a_{l_j m_k} S_{m_k}\right)}{\sum_{m_k} S_{m_k}}, & j \neq k, \\ R_{kk} = 4\pi - \sum_{j \neq k} R_{jk}, & j = k, \\ e_{ji} = \sum_{l_j} b_{l_j i}. & \end{cases} \qquad (13)$$

The summation is performed over the indices $l_j$, $m_k$, which number the small surface meshes in a large surface mesh $j(k)$. Here, $S_{l_j}$ is the surface area of small meshes, $a_{l_j m_k}$ and $b_{l_j i}$ are the matrix elements for small surface meshes calculated by formulas (9) and (7), respectively. For the elements $e_{ji}$, a normalization condition analogous to (8) is applicable:

$$\sum_j e_{ji} = -4\pi. \qquad (14)$$

Its use for error correction,

$$e_{ij} \rightarrow -4\pi \dfrac{e_{ij}}{\sum_{i=1}^{N} e_{ij}}$$

in the DISOLV program [4–6] gave a significant increase in accuracy.

# 3. DERIVATION OF THE COSMO EQUATION

The COSMO, a method of solving the electrostatic problem described at the beginning of the previous section, in which it is assumed that the dielectric constant inside the molecule is $\varepsilon_{in} = 1$ and $\varepsilon_{out} = \infty$ outside it. In this extreme case, the external environment can be considered a conductor, so the problem reduces to that of a cavity inside an infinite conductor, which certainly simplifies it. As in the previous section, we divide the surface into small meshes, located at points $\mathbf{r}_i$ and having a charge $q_i$. From the general considerations of electrostatics, the potentials of the surface and the outer region are set equal to zero to obtain the following
COSMO equation:

$$\sum_{k=1}^{N} \frac{q_k}{|\mathbf{r}_m - \mathbf{r}_k|} + \left( \sum_{i} \frac{Q_i}{|\mathbf{r}_m - \mathbf{R}_i|} \right) + \phi_{mm} = 0, \qquad (15)$$

where $\varphi_{mm}$ - is the potential of a surface mesh. In the matrix form it can be rewritten as

$$A_{(COSMO)} q = -B_{(COSMO)} Q, \qquad (16)$$

where $A_{(COSMO)}$ is a symmetric matrix with elements

$$a_{mk}^{(COSMO)} \approx |\mathbf{r}_m - \mathbf{r}_k|^{-1}, m \neq k, \qquad (17)$$

$$a_{mm}^{(COSMO)} = \frac{\phi_{mm}}{q_m},$$

whereas the elements of the matrix $B_{(COSMO)}$ are given by

$$b_{mj}^{(COSMO)} \approx |\mathbf{r}_m - \mathbf{R}_j|^{-1}. \qquad (18)$$

The potential of a small flat surface mesh is defined as [2, 4–6]:

$$\phi_{mm} \approx q_m \frac{2\sqrt{3.83}}{\sqrt{S_i}}. \qquad (19)$$

For small almost flat surface meshes, this expression gives a good approximation, as evidenced by calculations by the DISOLV program [4–6]. However, we are interested in a formula for the potential of a larger surface mesh. This formula should take into account its curvature and involve the parameters of the surface meshes themselves, similarly to expression (10). Let us obtain an exact normalization condition for the columns (rows) of the symmetric matrices and A$_{(COSMO)}$ and B$_{(COSMO)}$, which would allow us to determine the exact value of the diagonal elements. This problem reduces to the electrostatic mirror problem—finding the distribution of nonzero charge over the initial cavity filled with metal, with a vacuum outside, as we shall see below. This charge distribution is subjected to the following conditions:

(1) The electric field strength at the surface of the cavity (from inside) is zero.

(2) The electric field strength inside the cavity (including the points at which point charges were placed) is zero.

(3) The potential inside the cavity (including the points at which point charges were placed) is a constant $C_{val}$. The potential at infinity is taken as zero.

(4) The potential at the cavity surface is equal to the same constant, $-C_{val}$. The algorithm for solving this problem is described in detail in the Appendix.

### a. Коррекция ошибок дискретизации для матрицы $B^T_{(COSMO)}$ из условия III)

### 3.1. Correction of the Discretization Errors $B^T_{(COSMO)}$ for the Matrix Based on Condition (3)

Because of the discretization error, only an approximate equality holds:

$$B^T_{(COSMO)} q^{Norm} \approx 1_M C_{val}. \qquad (20)$$

, where $1_M$ is the column consisting of $M$ units, or in an expanded form,

$$\sum_m b_{(COSMO)mi} q_m^{Norm} \approx C_{val} \quad \forall i. \qquad (21)$$

wherefrom, the exact value of the constant $C_{val}$ is given by

$$C_{val} = \frac{1}{M} \sum_{i=1}^{M} \sum_m b_{(COSMO)mi} q_m^{Norm}. \qquad (22)$$

Correction of the discretization errors is accomplished according to

$$b_{(COSMO)mi} = b_{(COSMO)mi} \frac{C_{val}}{\sum_m b_{(COSMO)mi} q_m^{Norm}}. \tag{23}$$

Now, equalities (20) and (21) are satisfied exactly, not approximately.

### 3.2. Calculation of the Diagonal Elements of the Matrix $A_{(COSMO)}$.

Condition (4) can be written in the matrix form as

$$A_{(COSMO)} q^{Norm} = 1_N C_{val}, \tag{24}$$

where $1_N$ is the column of units. This condition makes it possible to determine the diagonal elements and, consequently, the potentials of the surface meshes:

$$a_{(COSMO)ii} \approx \frac{2\sqrt{3.83}}{\sqrt{S_i}} \quad, \text{если} \quad q_i^{Norm} << \frac{\sum q_j^{Norm}}{N}, \tag{25}$$

Otherwise,

$$\sum_{j \neq i} a_{(COSMO)ij} q_j^{Norm} + a_{(COSMO)ii} q_i^{Norm} = C_{val} \tag{26}$$

and

$$a_{(COSMO)ii} = \frac{C_{val} - \sum_{j \neq i} a_{(COSMO)ij} q_j^{Norm}}{q_i^{Norm}}. \tag{27}$$

### 3.3. Normalization Conditions for COSMO Surface Charges

To derive the normalization condition, we multiply both sides of (16) by $(q^{Norm})^T$:

$$(q^{Norm})^T A_{(COSMO)} q = -(q^{Norm})^T B_{(COSMO)} Q. \tag{28}$$

Then, (20) and (24) yield

$$1_N q = -1_M Q. \tag{29}$$

In the expanded form, it reads as:

$$\sum_{j=1}^{N} q_j = -\sum_{i=1}^{M} Q_i. \tag{30}$$

Using conditions (23) and (27) leads to a decrease in numerical calculation errors. In addition, it leads to automatic satisfaction of theoretical conditions (30).

### 3.4. Algorithm of Solution

The COSMO equation for the charges of the surface meshes was solved by an iterative scheme of the conjugate gradient method applied to linear equations defined by a symmetric positive definite matrix [7]. The conjugate gradient method minimizes the energy described by the following quadratic form:

$$\Delta G_{pol}(q_v) = \frac{1}{2} q_v^T A_{(COSMO)} q_v - f^T q_v, \tag{31}$$

Where $f = -B_{(COSMO)} Q$ is the column on the right-hand side of the COSMO equation, is the column of charges of surface meshes. The minimum of the quadratic form $q$ corresponds to the solution of the COSMO equation.
The initial iteration steps are given by

$$q^{(0)} = 0, \; p^{(0)} = f, \; r^{(0)} = f. \tag{32}$$

The iteration steps are specified by

$$\begin{cases} q^{(i+1)} = q^{(i)} + \alpha_i p^{(i)}, \\ q^{(i+1)} = q^{(i)} - \alpha_i A p^{(i)}, \quad \left(\alpha_i = \dfrac{(r^{(i)})^T r^{(i)}}{(p^{(i)})^T A p^{(i)}}\right), \\ p^{(i+1)} = r^{(i+1)} + \beta_i p^{(i)}, \quad \left(\beta_i = \dfrac{(r^{(i+1)})^T r^{(i+1)}}{(r^{(i)})^T r^{(i)}}\right), \end{cases}$$

(33)

where $q^{(i)}$ is the column of charges of surface meshes, whereas $p^{(i)}$ and $r^{(i)}$ are the columns of auxiliary variables. Using expression for the energy (31) at each iteration step, we obtain the error in the energy with respect to its value at the minimum, given by the column $q$:

$$\Delta G_{pol}^{(i)} = -\tfrac{1}{2}(q^{(i)})^T r^{(i)} + \tfrac{1}{2} q^T r^{(i)}. \tag{34}$$

The upper limit of its modulus reads as

$$\left|\Delta G_{pol}^{(i)}\right| < \tfrac{1}{2}\|r^{(i)}\|\left(\sum_j |Q_j|\right) + \tfrac{1}{2}\left|(q^{(i)})^T r^{(i)}\right|. \tag{35}$$

We used here the inequality $\sum_i |q_i| \le |Q_n| + |Q_p|$, where $Q_n$ and $Q_p$ are the sums of the negative and positive charges in the cavity, respectively. Indeed, any line of force emanating from a surface charge can only end at a charge inside the cavity, since the cavity surface and the outer region are equipotential; this condition is violated if the line of force ends at the cavity or leaves the outer region. Therefore, the value of positive components of the charge cavity does not exceed the absolute value of the sum of all negative charges in the cavity. A similar statement is true for negative components.

### 3.5. Derivatives of the Diagonal Elements of the Matrix $A_{(COSMO)}$.

Sometimes it becomes necessary to calculate the derivative of the diagonal elements of the matrix $A_{(COSMO)}$ with respect to some parameter α, for example, the average radius of the molecule, the coordinates of the atoms, etc. Direct differentiation of Eq. (27) is difficult, since the very charge distribution is a complicated function of α. However, the derivative of the diagonal elements can be found in another way:

$$a_{(COSMO)ii} \approx \frac{C_S}{\sqrt{S_i}}$$

$$\frac{\partial a_{(COSMO)ii}}{\partial \alpha} = -C_S \frac{\frac{\partial S_i}{\partial \alpha}}{2\sqrt{S_i^3}} \approx -a_{(COSMO)ii}\sqrt{S_i}\frac{\frac{\partial S_i}{\partial \alpha}}{2\sqrt{S_i^3}} = -\frac{a_{(COSMO)ii}}{2S_i}\frac{\partial S_i}{\partial \alpha} \quad (36)$$

Note that the final expression for the derivative does not contain the constant for which there are only rough estimates.

## 4. ENLARGED SURFACE MESHES FOR THE COSMO

### 4.1. Calculation of the Size of the Surface Meshes

Enlarged meshes are formed in the same way as in the Totorov–Abagyan method for enlarged surface meshes in the PCM [1]. Then, the COSMO equation is derived for enlarged surface meshes. This is performed identically as in the Totorov–Abagyan method [1]:
(1) Small surface meshes are combined into larger ones. The charge density on all small meshes belonging to a large one are set identical.
(2) New matrix elements for $L$ (the number of surface atoms) large surface meshes are determined:

$$Yq^{big}_{(COSMO)} = -TQ, \quad (37)$$

$$\begin{cases} y_{jk} = \frac{\left(\sum_{l_j}\sum_{m_k} a_{(COSMO)l_j m_k} S_{m_k}\right)}{\sum_{m_k} S_{m_k}} \quad j \neq k \\ t_{ji} = \sum_{l_j} b_{(COSMO)l_j i} \end{cases} \quad (38)$$

To determine the diagonal elements of the matrix $Y$ and to normalize the matrix $T$, it is necessary again to solve the problem of the distribution of nonzero charge $(q^{Norm})^{big}$. To solve this problem, it suffices to make use of the consequences from condition **2** the previous section: the forces at the points of placement of the charges are zeros. For this purpose, we again split large meshes into small ones, considering the surface charge density within the large surface meshes being constant:

$$\sum_j (q_j^{Norm})^{big} \sum_{l_j} \frac{1}{|\mathbf{R}_i - \mathbf{r}_{l_j}|^3} (\mathbf{R}_i - \mathbf{r}_{l_j}) = 0 \quad \forall i. \tag{39}$$

To solve Eq. (39), 1we use the Gram–Schmidt orthogonalization method to find $(q^{Norm})^{big}$, since the number of surface meshes is small. However, it is possible to use the iterative method described in the *Appendix2*. From condition **3**, we obtain

$$T^T (q^{Norm})^{big} = 1_M C_{val}, \tag{40}$$

where $1_M$ is the $M$ column of $M$ units. In the expanded form, Eq. (40) reads as

$$\sum_m t_{mi} (q_m^{Norm})^{big} \approx C_{val} \quad \forall i. \tag{41}$$

This equation holds approximately because of the discretization errors. Hence,

$$C_{val} = \frac{1}{M} \sum_{i=1}^{M} \sum_m t_{mi} (q_m^{Norm})^{big}. \tag{42}$$

The discretization error for the matrix $T$ is corrected using the formula

$$t_{mi} = t_{mi} \frac{C_{val}}{\sum_m t_{mi} (q_m^{Norm})^{big}}. \tag{43}$$

From condition **4**, we have

$$Y(q^{Norm})^{big} = 1_L C_{val}, \tag{44}$$

where $1_L$ is the column of units of length $L$. Then, from this condition, the diagonal elements of the matrix $Y$ and, accordingly, the proper potentials of the surface meshes are determined:

$$\sum_{j \neq i} y_{ij} (q_j^{Norm})^{big} + y_{ii} (q_i^{Norm})^{big} = C_{val}, \tag{45}$$

whence the diagonal elements are given by

$$y_{ii} = \frac{C_{val} - \sum_{j \neq i} y_{ij}(q_j^{Norm})^{big}}{(q_i^{Norm})^{big}}. \tag{46}$$

Multiplying both sides of Eq. (37) by $\left((q^{Norm})^{big}\right)^T$, yields

$$\left((q^{Norm})^{big}\right)^T Y q_{(COSMO)}^{big} = -\left((q^{Norm})^{big}\right)^T TQ, \tag{47}$$

The normalization condition for the charges reads as:

$$(1_L)^T q_{(COSMO)}^{big} = -(1_M)^T Q. \tag{48}$$

or in expanded form:

$$\sum_{j=1}^{L} q_{(COSMO)j}^{big} = -\sum_{i=1}^{M} Q_i. \tag{49}$$

### 4.2 Derivatives of the Diagonal Elements of the Matrix Y

Let the matrix $T_{surf}$ for those numbers $i$ that refer to the surface atoms be

$$t_{(surf)ij} = t_{ij}. \tag{50}$$

The derivatives of the normalizing surface charges with respect to some parameter $\dfrac{\partial (q^{Norm})^{big}}{\partial \alpha}$ are obtained from Eq. (40):

$$T_{surf}^T \frac{\partial (q^{Norm})^{big}}{\partial \alpha} = -\frac{\partial T_{surf}^T}{\partial \alpha}(q^{Norm})^{big}. \tag{51}$$

The derivative of the diagonal elements of the matrix Y (formula (44)) is expressed as $\sum_{j \neq i} \dfrac{\partial y_{ij}}{\partial \alpha}(q_j^{Norm})^{big} + \dfrac{\partial y_{ii}}{\partial \alpha}(q_i^{Norm})^{big} = -\sum_{j \neq i} y_{ij}\dfrac{\partial (q_j^{Norm})^{big}}{\partial \alpha} - y_{ii}\dfrac{\partial (q_i^{Norm})^{big}}{\partial \alpha}.$ (52)

Hence, we obtain

$$\frac{\partial y_{ii}}{\partial \alpha} = -\frac{\sum_{j \neq i} y_{ij} \frac{\partial (q_j^{Norm})^{big}}{\partial \alpha} + y_{ii} \frac{\partial (q_i^{Norm})^{big}}{\partial \alpha} + \sum_{j \neq i} \frac{\partial y_{ij}}{\partial \alpha} (q_j^{Norm})^{big}}{(q_i^{Norm})^{big}}. \qquad (53)$$

where the derivatives of the charges are given by (51).

## 5. CONCLUSIONS

In this work, we obtained normalization condition for the columns (rows) of the COSMO matrix, which then were used to derive exact expressions for the diagonal elements of the main COSMO matrix. For these purpose, it is necessary to solve the electrostatic problem of charge distribution over the metal surface. A method of enlarged surface meshes for the COSMO, similar to that used in the PCM [4], was developed. This method makes it possible to approximately solve the COSMO equation and determine the Born radii for the SGB model [8].

# APPENDIX

## *1. Algorithm for Calculating the Normals from the Coordinates of the Surface Meshes*

In the COSMO, the normals are usually not specified—only the coordinates of the surface meshes are given. In this case, the normal to a mesh can be determine by the following method. The normal to the mesh is sought for as the average of the normals to the triangular meshes with the vertex at the center the polygonal mesh weights proportional to the squares of the surface area of the triangles:

$$a_{gi} = r_{gi} - r_g, \tag{54}$$

where $r_g$ is the radius-vector of a given centers, $r_{gi}$ are the radii-vectors of the adjacent surface meshes, and $i=1,...,N_{eg}$ is the number of adjacent surface meshes. Let us introduce the notation

$$a_{gi\_next} = \begin{cases} a_{g(i+1)} & \text{for } 1 \leq i < N_{eg} \\ a_{g1} & \text{for } i = N_{eg} \end{cases}. \tag{55}$$

In the first step of the algorithm,

$$n_g = \frac{\sum_{i=1}^{N_{eg}} ([a_{gi} \times a_{gi\_next}])}{\left|\sum_{i=1}^{N_{eg}} ([a_{gi} \times a_{gi\_next}])\right|}$$

with the normal to the current surface mesh is given by

$$n_g = \frac{\sum_{i=1}^{N_{eg}} ([a_{gi} \times a_{gi\_next}] sign(n_{near}[a_{gi} \times a_{gi\_next}]))}{\left|\sum_{i=1}^{N_{eg}} ([a_{gi} \times a_{gi\_next}] sign(n_{near}[a_{gi} \times a_{gi\_next}]))\right|} \tag{56}$$

where $n_{near}$ - normal of neighboring surface element.

In this case, all the normals will have the same direction, either inward or outward. However, we need that they be directed outward. Then, the volume of the surface is calculated by the formula

$$V = \frac{1}{3}\oint_S (\mathbf{r} \cdot \mathbf{n}) dS \tag{57}$$

where **r** is the radius vector of the current point at the surface and **n** is the normal to it. If $V < 0$, signs of the normals should be changed:

$$\mathbf{n}_g = -\mathbf{n}_g. \tag{58}$$

## 2. Algorithm for Determining the Distribution of Nonzero Charge over the Initial Cavity with Metal inside and Vacuum outside It

Let us describe the algorithm for solving the mirror electrostatic problem of determining the distribution of charge over the initial cavity filled with metal and vacuum outside of it. Conditions **1**–**4** to be met by such a distribution of charges are given in Section 3.

Then, the equations defined by conditions **1** and **2** can be written as the condition of orthogonality of the vector of charges to the known system of vectors. According to condition **1**, the directions *perpendicular* to the normal of the surface mesh is given by

$$\left[ \sum_{i \neq j} q_j^{Norm} \frac{(\mathbf{r}_i - \mathbf{r}_j)}{|\mathbf{r}_i - \mathbf{r}_j|^3} \times \mathbf{n}_i \right] = 0. \tag{59}$$

The condition for the directions along the normals to the surface elements of the equation read as

$$\left( \frac{\partial \varphi_{in}(\mathbf{r})}{\partial n} \right) = 0. \tag{60}$$

Upon dividing into surface meshes, Eq. (2), subjected to conditions (60) and $Q = 0$, can be written in the matrix form as follows (by analogy with derivation of Eqs. (6) and (5)):

$$-Aq^{Norm} + 4\pi q^{Norm} = 0. \tag{61}$$

The same can be written in terms of the matrix elements $a_{ij}$:

$$-\sum_{j \neq i} a_{ij} q_j^{Norm} - a_{ii} q_i^{Norm} + 4\pi q_i^{Norm} = 0 \tag{62}$$

or in the expanded vector form

$$-n_i S_i \sum_{j \neq i} q_j^{Norm} \frac{(r_i - r_j)}{|r_i - r_j|^3} + q_i^{Norm} \left( \sum_{j \neq i} \frac{(r_j - r_i)}{|r_j - r_i|^3} n_j S_j \right) = 0. \tag{63}$$

For flat small meshes, this reduces to the equation

$$-n_i S_i \sum_{j \neq i} q_j^{Norm} \frac{(r_i - r_j)}{|r_i - r_j|^3} + q_i^{Norm} 2\pi = 0. \tag{64}$$

From condition **2**, for the points of location of the centers of the atoms, we obtain

$$\sum_m \frac{q_m^{Norm}}{|R_i - r_m|^3} (R_i - r_m) = 0 \quad \forall i. \tag{65}$$

Let us use an iterative method for calculating from Eqs. (58), (62), and (64). This is, in fact, the procedure for determining a vector orthogonal to a given set of vectors.

(1) As an initial approximation, we choose the vector $q^{(0)Norm} = S$, where $S$ is the column consisting of the areas of the surface meshes involved, and normalize it by unity: $\|q^{(0)Norm}\| = 1$

(2) From the equipotential surface condition, all the charges of the surface meshes have the same sign, which we choose positive $q^{norm} > 0$. Hence, $(q^{(0)Norm})^T \cdot q^{Norm} > 0$, and therefore, the initial approximation has a nonzero projection on the desired vector, which makes it suitable for iterations.

(3) The Gram–Schmidt orthogonalization procedure [9] seems too costly; therefore, we used an iterative search process:

A) Let us normalize all 3N +3 M vectors in (58), (62) and (64) by unity and denote this set of vectors as $d_k$, where $k=1,...,3N+3M$; $\|d_k\|=1$. The current vector will be successively permutated with each vector from the set

$$q_j^{(k)Norm} = q_j^{(k-1)Norm} - ((q_j^{(k-1)Norm})^T d_k) q_j^{(k-1)Norm}, \quad k = 1,...,3N + 3M.$$

If a component of this vector turns out to be negative, it is set equal to zero $(q_j^{(k)Norm})_m = 0$.

B) If $\| q_j^{(3N+3M)Norm} \| \approx 1$, the iteration procedure is stopped, if not, return to Step A with $q_{j+1}^{(0)Norm} = q_j^{(3N+3M)Norm} / \| q_j^{(3N+3M)Norm} \|$ is made.

# УКРУПНЕННЫЕ ПОВЕРХНОСТНЫЕ ЭЛЕМЕНТЫ И УСЛОВИЯ НОРМИРОВКИ ДЛЯ СТОЛБЦОВ И СТРОК МАТРИЦ МЕТОДА COSMO.

Купервассер* О.Ю.[1], Кикоть** И.П.[2]

[1]*Научно-исследовательский вычислительный центр Московского государственного университета им.М.В. Ломоносова*

[2]*Учреждение Российской академии наук Институт химической физики им. Н.Н.Семенова Российской академии наук, Москва*

*<u>E-mail:</u> olegkup@yahoo.com,

**<u>E-mail:</u> irakikotx@gmail.com

В более ранней работе Тотрова и Абагяна были найдены условия нормировки столбцов матрицы PCM и разработан метод укрупненных поверхностных элементов. Мы разработали аналогичные методы для метода COSMO. Эти методы позволят ввести более крупные поверхностные элементы без потери точности, а также позволят находить расчетно-быстрые приближенные значения энергии сольватации и радиусов Борна для метода SGB (*Surface Generalized Born*). Кроме того, предложенные в данной работе корректировки численных результатов позволили значительно увеличить точность вычислений.

Ключевые слова: электростатика, энергия сольватации, укрупненные поверхностные элементы, диагональные члены, COSMO.

## 1. Введение.

Хорошо известно, что адекватный учет взаимодействия заряженной молекулы и окружающего ее растворителя (чаще всего воды) очень важен при моделировании различных химических систем, в частности, биомолекул. Существует несколько различных подходов к моделированию растворителя. В данной работе мы остановим внимание на моделях, в которых растворитель рассматривается как сплошная среда с однородной диэлектрической проницаемостью, а молекула – как область внутри этой среды также с однородной диэлектрической проницаемостью, причем на границе области диэлектрическая проницаемость терпит разрыв. При таком подходе решение электростатической задачи сводится к нахождению поверхностных зарядов, индуцированных внесением заряженной молекулы в диэлектрик. Ниже приводится краткое описание двух таких моделей – PCM [1] и COSMO [3].

В работе рассматривается условия нормализации для столбцов (строк) матриц COSMO (*COnductor-like Screening MOdel*) [3]. Эти методы позволяют найти точные значения диагональных элементов симметричной матрицы COSMO и обеспечить выполнение условия, связывающего заряд внутри полости с поверхностным зарядом. Также рассмотренные условия позволяют разработать приближенный метод укрупненных поверхностных элементов для COSMO, аналогичный методу укрупненных поверхностных элементов для PCM (*Polarized Continuum Model*) [1,4]. Этот метод должен давать лучший результат, чем в случае PCM, поскольку его матричные элементы зависят лишь от координат поверхностных элементов и менее чувствительны к ошибкам укрупнения размеров поверхностных элементов.

Статья построена следующим образом. В первой части нашей работы мы даем краткое описание условий нормировки матриц PCM и метод укрупненных поверхностных элементов, разработанный в [1]. Во второй части мы определяем условия нормировки для матриц COSMO и находим формулы для диагональных элементов симметричной матрицы. В третьей части мы определяем метод укрупненных элементов для COSMO.

## 2. PCM-метод с укрупненными поверхностными элементами для $\varepsilon=\infty$.

Метод PCM [1,4] (модель поляризуемой диэлектрической среды) представляет собой метод нахождения точного решения электростатической задачи, в которой точечные заряды, соответствующие заряженным атомам молекулы растворяемого вещества, окружены непрерывным диэлектриком с заданной диэлектрической проницаемостью $\varepsilon$. В этом случае электрическое поле в диэлектрической среде будет равно электрическому полю, которое создавалось бы в вакууме зарядами внутри и поверхностными зарядами, индуцированными молекулой растворяемого вещества на поверхности, разделяющей диэлектрик-растворитель и молекулу. Таким образом, вместо решения уравнения Пуассона в трехмерном пространстве необходимо найти поверхностные заряды.

Рассмотрим точную постановку задачи о взаимодействии точечных зарядов внутри диэлектрической полости с диэлектриком. Пусть имеются две односвязные области пространства, разделённые замкнутой поверхностью $\Xi$. Внутренняя область $\Omega_{in}$ имеет диэлектрическую проницаемость $\varepsilon_{in}=1$, а внешняя область $\Omega_{ex}$ – проницаемость $\varepsilon=\infty$. Стоит отметить, что метод PCM вполне успешно может быть применен и в случае конечной диэлектрической проницаемости внешней среды. Однако для единообразия с дальнейшим описанием изложим его в предельном случае $\varepsilon=\infty$. Во внутренней части расположена система зарядов $Q_i$, находящихся в точках $\boldsymbol{R}_i$. В результате взаимодействия зарядов с диэлектриком на границе раздела областей индуцируется поверхностный заряд плотности $\sigma$. Электростатический потенциал может быть выражен как сумма потенциалов точечных зарядов внутри и поверхностных зарядов:

$$\varphi(r) = \sum_i \frac{Q_i}{\varepsilon_{in}|r-R_i|} + \int_\Xi \frac{\sigma(r')}{|r-r'|} dS'. \qquad (1)$$

Производная вдоль нормали на поверхности с внутренней стороны

$$\left(\frac{\partial \varphi_{in}(r)}{\partial n}\right) = \left(-\int_\Xi \frac{\sigma(r')((r-r')\cdot n(r))}{|r-r'|^3} dS' + 2\pi\sigma\right) - \sum_i \frac{Q_i((r-R_i)\cdot n(r))}{\varepsilon_{in}|r-R_i|^3}. \qquad (2)$$

Поверхностный интеграл является сингулярным при $r=r'$, в этом случае определим его следующим образом:

$$\int_\Xi \frac{\sigma(r')((r-r')\cdot n(r))}{|r-r'|^3} dS' = \lim_{\delta\to 0} \int_{\Xi/(|r-r'|<\delta)} \frac{\sigma(r')((r-r')\cdot n(r))}{|r-r'|^3} dS'. \qquad (3)$$

Из теоремы Остроградского - Гаусса

$$4\pi\sigma = \frac{\partial \varphi_{in}(r)}{\partial n}. \qquad (4)$$

Из уравнений (2) и (4) получаем интегральное уравнение для σ:

$$4\pi\sigma(r) = \left(-\int_\Xi \frac{\sigma(r')((r-r')\cdot n(r))}{|r-r'|^3} dS' + 2\pi\sigma(r)\right) - \sum_j \frac{Q_j((r-R_j)\cdot n(r))}{\varepsilon_{in}|r-R_j|^3}. \qquad (5)$$

Для численного решения этого интегрального уравнения введем дискретизацию – разобьем поверхность на *N* малых поверхностных элементов, которые можно считать точечными зарядами. Пусть $q_i$ – заряд *i*-ого поверхностного элемента, $q=\{q_i\}_{i=1,N}$ – вектор-столбец зарядов поверхностных элементов, $S_i$ – площадь *i*-ого поверхностного элемента, $r_i$ - вектор, определяющий положение центра *i*-ого поверхностного элемента, $n_i$ - вектор нормали к *i*-ому поверхностному элементу, проведенный через его центр и направленный в сторону диэлектрика-растворителя. Число точечных зарядов внутри молекулы обозначим за *M*, $Q=\{Q_j\}_{j=1,M}$ – вектор-столбец зарядов атомов молекулы. Тогда уравнение (5) может быть записано в матричной форме:

$$4\pi q = ((-Aq + 2\pi q) + 2\pi q) + BQ,$$

Здесь первый интеграл для удобства записывается в виде $(-Aq+2\pi q)$. Тогда после сокращения одинаковых слагаемых получаем уравнение

$$Aq = BQ, \qquad (6)$$

где $A=\{a_{ij}\}$ – квадратная матрица размера $N$x$N$, $B=\{b_{ij}\}$ – матрица размера $N$x$M$, зависящая от геометрических параметров поверхностных элементов и положения зарядов $Q_i$ атомов молекулы. Матричные элементы матрицы $B$ имеют вид:

$$b_{ij} = -\frac{\mathbf{n}_i \cdot (\mathbf{r}_i - \mathbf{R}_j)}{|\mathbf{r}_i - \mathbf{R}_j|^3} S_i. \qquad (7)$$

Для столбцов матрицы $B$ должно выполняться следующее тождество:

$$\sum_{i=1}^{N} b_{ij} = -4\pi. \qquad (8)$$

Это тождество может быть использовано для коррекции численных ошибок дискретизации, возникающие при разбиении на конечные поверхностные элементы:

$$b_{ij} \to -4\pi \frac{b_{ij}}{\sum_{i=1}^{N} b_{ij}}.$$

Использование такой коррекции в программе DISOLV [5-7] дало значительное увеличение точности. Матричные элементы матрицы $A$ имеют следующий вид [1,4]:

$$a_{ij} = \begin{cases} \dfrac{\mathbf{n}_i (\mathbf{r}_i - \mathbf{r}_j)}{|\mathbf{r}_i - \mathbf{r}_j|^3} S_i, & i \neq j, \\ a_{jj} = 4\pi - \sum_{i \neq j} a_{ij}, & i = j. \end{cases} \qquad (9)$$

Геометрический смысл $2\pi - \sum_{i \neq j} a_{ij}$, – поправка на кривизну поверхности ($a_{jj}=2\pi$ для плоского малого элемента). Из выражения (9) следует условие нормировки для столбцов матрицы A:

$$\sum_{i} a_{ij} = 4\pi. \qquad (10)$$

Из уравнения (6) путем суммирования всех строк, а затем использованием тождеств (8) и (10), получаем тождество, связывающее суммарный поверхностный заряд с суммарным зарядом внутри полости:

$$\sum_{j=1}^{N} q_j = -\sum_{i=1}^{M} Q_i . \qquad (11)$$

Для убыстрения PCM может быть использован метод Тотрова и Абагяна с увеличенными поверхностными элементами [1]. Он описывается теми исходными интегральными уравнениями (6-10), что и обычная PCM модель, но при дискретизации поверхностные элементы берутся существенно более крупными. Введем эти укрупненные поверхностные элементы следующим образом. Рассмотрим все поверхностные атомы. Объединим все малые поверхностные элементы в группы, состоящие из поверхностных элементов, наиболее близких к данному поверхностному атому.

Будем считать, что поверхностная плотность заряда одинакова во всех точках для любого укрупненного поверхностного элемента. Тогда уравнение PCM с укрупненными поверхностными элементами в матричной форме можно записать в следующем виде:

$$Rq^{big} = EQ , \qquad (12)$$

где $q^{big}$ - столбец зарядов укрупненных поверхностных элементов.
$E=\{e_{ij}\}$ – матрица размера $N_{\{surf\}} \times M$, $R=\{R_{ij}\}$ – квадратная матрица размера $N_{\{surf\}} \times N_{\{surf\}}$. $N_{\{surf\}}$ – число укрупненных поверхностных элементов.
Матричные элементы для укрупненных элементов будут вычисляться по формулам:

$$\begin{cases} R_{jk} = \dfrac{\left(\sum\limits_{l_j}\sum\limits_{m_k} a_{l_j m_k} S_{m_k}\right)}{\sum\limits_{m_k} S_{m_k}}, & j \neq k, \\ R_{kk} = 4\pi - \sum\limits_{j \neq k} R_{jk}, & j = k, \\ e_{ji} = \sum\limits_{l_j} b_{l_j i}. & \end{cases} \qquad (13)$$

Суммирование здесь проводится по индексам $l_j$, $m_k$, которые соответствуют всем малым поверхностным элементам, попавшим в один крупный поверхностный элемент с номером $j(k)$. Здесь $S_{l_j}$ - площадь малых поверхностных элементов, $a_{l_j m_k}$ - матричные элементы для малых поверхностных элементов (9), а матричные элементы $b_{l_j i}$ для малых поверхностных элементов вычисляются по формуле (7).

Для элементов $e_{ji}$ выполняется условие нормализации, аналогичное (8)

$$\sum_j e_{ji} = -4\pi. \qquad (14)$$

Использование его для коррекции ошибок

$$e_{ij} \to -4\pi \dfrac{e_{ij}}{\sum\limits_{i=1}^{N} e_{ij}}$$

в программе DISOLV [5-7] дало значительное увеличение точности.

### 3. **Вывод уравнения COSMO**

Метод COSMO (модель экранирования, подобного экранированию в проводниках) − метод решения электростатической задачи, описанной в начале предыдущего раздела, в которой предполагается, что диэлектрическая проницаемость внутри молекулы равна 1, а диэлектрическая проницаемость внешней среды бесконечна: $\varepsilon_{out}=\infty$. В этом предельном случае внешнюю среду можно считать проводником; таким образом, задача сводится к рассмотрению полости внутри бесконечного проводника, что, разумеется, упрощает ее.

Как и в предыдущем разделе, разобьем поверхность на малые элементы, расположение в точках $r_i$ и имеющие заряд $q_i$. Из общих соображений электростатики потенциал поверхности и внешней области одинаков и принимается за ноль. Отсюда получаем уравнение COSMO [3]:

$$\sum_{k=1}^{N}\frac{q_k}{|r_m - r_k|} + \left(\sum_{i}\frac{Q_i}{|r_m - R_i|}\right) + \phi_{mm} = 0, \qquad (15)$$

где $\varphi_{mm}$ - собственный потенциал поверхностного элемента. В матричной форме его можно переписать так:

$$A_{(COSMO)}q = -B_{(COSMO)}Q, \qquad (16)$$

где $A_{(COSMO)}$ - симметричная матрица с элементами

$$a_{mk}^{(COSMO)} \approx |r_m - r_k|^{-1}, m \neq k, \qquad (17)$$

$$a_{mm}^{(COSMO)} = \frac{\phi_{mm}}{q_m},$$

а элементы матрицы $B_{(COSMO)}$ задаются следующим выражением:

$$b_{mj}^{(COSMO)} \approx |r_m - R_j|^{-1}. \qquad (18)$$

Для малого плоского поверхностного элемента собственный потенциал выражается следующей формулой [3,5-7]:

$$\phi_{mm} \approx q_m \frac{2\sqrt{3.83}}{\sqrt{S_i}} \qquad . \qquad (19)$$

Для малых почти плоских поверхностных элементов это выражение дает хорошее приближение, как свидетельствуют расчеты с помощью программы DISOLV [5-7].

Мы, однако, заинтересованы в более точном выражении, применимом для собственного потенциала более крупного поверхностного элемента. Это выражение должно учитывать его кривизну и выражаться через параметры самих поверхностных элементов аналогично формуле (10).

Найдем точное условие нормировки для столбцов (строк) симметричной матрицы $A_{(COSMO)}$ и $B_{(COSMO)}$, которое позволит нам получить точное значение диагональных поверхностных элементов. Эта задача сводится к зеркальной электростатической задаче - нахождению распределения ненулевого заряда $q_m^{Norm}$ по исходной полости, заполненной металлом, с вакуумом вовне, как мы увидим ниже. Такое распределение заряда находится из следующих условий:

I) Напряженность электрического поля на поверхности полости (с внутренней стороны) равна нулю.

II) Напряженность электрического поля внутри полости (в том числе в точках, где были размещены точечные заряды) равна нулю.

III) Потенциал внутри полости (в том числе в точках, где были размещены точечные заряды) равен константе. Потенциал на бесконечности принимаем за ноль.

IV) Потенциал на поверхности полости равен той же константе.

Алгоритм решения этой задачи подробно изложен в Приложении.

## 3.1 Коррекция ошибок дискретизации для матрицы $B_{(COSMO)}^T$ из условия III)

Из-за ошибок дискретизации имеем лишь приближенное равенство:

$$B_{(COSMO)}^T q^{Norm} \approx 1_M C_{val}. \tag{20}$$

, где $1_M$ – столбец из M единиц

В развернутом виде:

$$\sum_m b_{(COSMO)mi} q_m^{Norm} \approx C_{val} \quad \forall i. \tag{21}$$

Отсюда определяем точное значение константы $C_{val}$:

$$C_{val} = \frac{1}{M} \sum_{i=1}^{M} \sum_m b_{(COSMO)mi} q_m^{Norm}. \tag{22}$$

Коррекция ошибок дискретизации для матрицы $B_{(COSMO)}$ производится следующим образом:

$$b_{(COSMO)mi} = b_{(COSMO)mi} \frac{C_{val}}{\sum_m b_{(COSMO)mi} q_m^{Norm}}. \tag{23}$$

Теперь равенства (20-21) выполняются точно, а не приближенно.

### 3.2 Подсчет диагональных элементов матрицы $A_{(COSMO)}$.

Условие IV) можно записать в матричной форме:

$$A_{(COSMO)} q^{Norm} = 1_N C_{val}, \tag{24}$$

где $1_N$ - столбец из $N$ единиц. Тогда из этого условия можно найти диагональные элементы матрицы $A$ и, соответственно, собственные потенциалы поверхностных элементов:

$$a_{(COSMO)ii} \approx \frac{2\sqrt{3.83}}{\sqrt{S_i}} \quad, \text{если} \quad q_i^{Norm} \ll \frac{\sum_j q_j^{Norm}}{N}, \tag{25}$$

Иначе

$$\sum_{j \neq i} a_{(COSMO)ij} q_j^{Norm} + a_{(COSMO)ii} q_i^{Norm} = C_{val} \tag{26}$$

и

$$a_{(COSMO)ii} = \frac{C_{val} - \sum_{j \neq i} a_{(COSMO)ij} q_j^{Norm}}{q_i^{Norm}}. \tag{27}$$

### 3.3 Условие нормализации поверхностных зарядов COSMO.

Домножая обе части уравнения (16) на $(q^{Norm})^T$, получаем:

$$(q^{Norm})^T A_{(COSMO)} q = -(q^{Norm})^T B_{(COSMO)} Q. \tag{28}$$

Следовательно, из (20) и (24)

$$1_N q = -1_M Q. \tag{29}$$

В развернутом виде:

$$\sum_{j=1}^{N} q_j = -\sum_{i=1}^{M} Q_i. \tag{30}$$

Использование (23) и (27) приводит к уменьшению численных ошибок. Кроме того, оно ведет к автоматическому выполнению теоретического условия на суммарный заряд поверхности (30).

### 3.4 Алгоритм решения

Уравнение COSMO для зарядов поверхностных элементов решалось с помощью итерационной схемы метода сопряженных градиентов, используемого для линейных уравнений, определяемых симметричной положительной определенной матрицей [8]. Метод сопряженных элементов минимизирует энергию, описываемую следующей квадратичной формой

$$\Delta G_{pol}(q_v) = \frac{1}{2} q_v^T A_{(COSMO)} q_v - f^T q_v, \tag{31}$$

где $f = -B_{(COSMO)}Q$ - столбец правой части уравнения COSMO, $q_v$ – столбец зарядов поверхностных элементов. Минимум квадратичной формы $q$ соответствует решению уравнения COSMO.

Начальные шаги итерации выбираются так:

$$q^{(0)} = 0, \ p^{(0)} = f, \ r^{(0)} = f. \tag{32}$$

Шаги итерации:

$$\begin{cases} q^{(i+1)} = q^{(i)} + \alpha_i p^{(i)}, \\ q^{(i+1)} = q^{(i)} - \alpha_i A p^{(i)}, \ \left(\alpha_i = \frac{(r^{(i)})^T r^{(i)}}{(p^{(i)})^T A p^{(i)}}\right), \\ p^{(i+1)} = r^{(i+1)} + \beta_i p^{(i)}, \ \left(\beta_i = \frac{(r^{(i+1)})^T r^{(i+1)}}{(r^{(i)})^T r^{(i)}}\right), \end{cases} \tag{33}$$

где $q^{(i)}$ – столбец зарядов поверхностных элементов, $p^{(i)}$ и $r^{(i)}$ – столбцы вспомогательных переменных.

Используя выражение для энергии (31) на каждом шаге итерации, получаем ошибку энергии по отношению к ее значению в минимуме, даваемым столбцом $q$:

$$\Delta G_{pol}^{(i)} = -\tfrac{1}{2}(q^{(i)})^T r^{(i)} + \tfrac{1}{2} q^T r^{(i)}. \tag{34}$$

Верхняя граница ее модуля определяется по формуле:

$$\left|\Delta G_{pol}^{(i)}\right| < \tfrac{1}{2}\left\|r^{(i)}\right\|\left(\sum_j |Q_j|\right) + \tfrac{1}{2}\left|(q^{(i)})^T r^{(i)}\right|. \tag{35}$$

Мы использовали здесь следующее неравенство: $\sum_i |q_i| \leq |Q_n| + |Q_p|$, где $Q_n$ и $Q_p$ – соответственно, сумма отрицательных и положительных зарядов в полости. Действительно, любая силовая линия, выходящая из заряда на поверхности, может закончиться только на заряде внутри полости, поскольку поверхность полости и внешняя область эквипотенциальны; это условие нарушится, если силовая линия закончиться на полости или уйдет во внешнюю область. Поэтому величина положительной компоненты заряда полости не превышает абсолютную величину суммы всех отрицательных зарядов в полости. Аналогичное утверждение верно и для отрицательной компоненты.

### 3.5 Производные диагональных элементов матрицы $A_{(COSMO)}$.

Часто возникает потребность посчитать производную диагональных элементов матрицы $A_{(COSMO)}$ по какому-нибудь параметру α, например, среднему радиусу молекулы, координатам атомов и т.д. Прямое дифференцирование выражения (27) весьма сложно, поскольку само распределение поверхностных зарядов $q^{Norm}$ сложным образом зависит от параметра $\alpha$. Однако производную диагональных элементов матрицы $A_{(COSMO)}$ можно найти по-другому:

$$a_{(COSMO)ii} \approx \frac{C_S}{\sqrt{S_i}}$$

$$\frac{\partial a_{(COSMO)ii}}{\partial \alpha} = -C_S \frac{\frac{\partial S_i}{\partial \alpha}}{2\sqrt{S_i^3}} \approx -a_{(COSMO)ii} \sqrt{S_i} \frac{\frac{\partial S_i}{\partial \alpha}}{2\sqrt{S_i^3}} = -\frac{a_{(COSMO)ii}}{2S_i} \frac{\partial S_i}{\partial \alpha}$$ (36)

Отметим, что в полученное выражение не входит константа $C_s$, для которой имеются только приближенные оценки.

## 4. Укрупненные поверхностные элементы для COSMO
### 4.1 Расчет величины поверхностных элементов

Укрупненные элементы формируются так же, как в методе Тоторова и Абагяна для укрупненных поверхностных элементов PCM[1]. Затем необходимо найти уравнение COSMO укрупненных поверхностных элементов. Делаем это совершенно аналогично методу Тоторова и Абагяна [1] для укрупненных поверхностных элементов PCM:

1) Объединяем малые поверхностные элемента в более крупные. Плотность заряда на всех малых элементах, относящихся к одному крупному, считаем одинаковой

2) Находим новые матричные элементы для L (число поверхностных атомов) больших поверхностных элементов:

$$Yq_{(COSMO)}^{big} = -TQ,$$ (37)

$$\begin{cases} y_{jk} = \frac{\left(\sum_{l_j} \sum_{m_k} a_{(COSMO)l_j m_k} S_{m_k}\right)}{\sum_{m_k} S_{m_k}} \quad j \neq k \\ t_{ji} = \sum_{l_j} b_{(COSMO)l_j i} \end{cases}$$ (38)

С целью нахождения диагональных элементов матрицы $Y$ и нормализации матрицы $T$ снова нужно решить задачу о распределении ненулевого заряда $(q^{Norm})^{big}$. Для решения этой задачи вполне достаточно следствий из условия II) предыдущего пункта - силы в точках, где были размещены точечные заряды, равны нулю. При его реализации мы снова разбиваем большие элементы на малые, считая поверхностную плотность заряда в границах больших поверхностных элементов постоянной:

$$\sum_j (q_j^{Norm})^{big} \sum_{l_j} \frac{1}{\left|\boldsymbol{R}_i - \boldsymbol{r}_{l_j}\right|^3} (\boldsymbol{R}_i - \boldsymbol{r}_{l_j}) = 0 \quad \forall i. \tag{39}$$

Для решения уравнений (39) уже применим метод ортогонализации Грамма - Шмидта для нахождения $(q^{Norm})^{big}$, поскольку число поверхностных элементов невелико. Можно, однако, использовать и итерационный метод из Приложения.

Из условия III) получаем:

$$T^T (q^{Norm})^{big} = 1_M C_{val}, \tag{40}$$

где $1_M$ – вектор единиц длиной M. В развернутом виде уравнение (40) выглядит следующим образом:

$$\sum_m t_{mi} (q_m^{Norm})^{big} \approx C_{val} \quad \forall i. \tag{41}$$

Это уравнение выполняется приближенно из-за ошибок дискретизации. Отсюда

$$C_{val} = \frac{1}{M} \sum_{i=1}^{M} \sum_m t_{mi} (q_m^{Norm})^{big}. \tag{42}$$

Коррекция ошибок дискретизации для матрицы $T$ производится по формуле

$$t_{mi} = t_{mi} \frac{C_{val}}{\sum_m t_{mi} (q_m^{Norm})^{big}}. \tag{43}$$

Из условия IV) имеем:

$$Y (q^{Norm})^{big} = 1_L C_{val}, \tag{44}$$

где $1_L$ - столбец из единиц длиной L. Тогда из этого условия можно найти диагональные элементы матрицы $Y$ и, соответственно, собственные потенциалы поверхностных элементов:

$$\sum_{j \neq i} y_{ij} (q_j^{Norm})^{big} + y_{ii} (q_i^{Norm})^{big} = C_{val} \ , \tag{45}$$

откуда для диагональных элементов

$$y_{ii} = \frac{C_{val} - \sum_{j \neq i} y_{ij} (q_j^{Norm})^{big}}{(q_i^{Norm})^{big}} \ . \tag{46}$$

Домножая обе части уравнения (37) на $\left((q^{Norm})^{big}\right)^T$, получаем:

$$\left((q^{Norm})^{big}\right)^T Y q_{(COSMO)}^{big} = -\left((q^{Norm})^{big}\right)^T TQ \ , \tag{47}$$

Получим условие нормализации для зарядов

$$(I_L)^T q_{(COSMO)}^{big} = -(I_M)^T Q \ . \tag{48}$$

Это же можно записать в развёрнутом виде:

$$\sum_{j=1}^{L} q_{(COSMO)j}^{big} = -\sum_{i=1}^{M} Q_i \ . \tag{49}$$

### 4.2 Производные диагональных элементов матрицы Y.

Введём матрицу $T_{surf}$ для тех номеров i, которые относятся к поверхностным атомам:

$$t_{(surf)ij} = t_{ij} \ . \tag{50}$$

Производные нормирующих поверхностных зарядов по какому-либо параметру $\dfrac{\partial (q^{Norm})^{big}}{\partial \alpha}$ находятся из уравнения (40):

$$T_{surf}^T \frac{\partial (q^{Norm})^{big}}{\partial \alpha} = -\frac{\partial T_{surf}^T}{\partial \alpha} (q^{Norm})^{big} \ . \tag{51}$$

Производная диагональных элементов матрицы $Y$ (из (44)) :

$$\sum_{j \neq i} \frac{\partial y_{ij}}{\partial \alpha} (q_j^{Norm})^{big} + \frac{\partial y_{ii}}{\partial \alpha} (q_i^{Norm})^{big} = -\sum_{j \neq i} y_{ij} \frac{\partial (q_j^{Norm})^{big}}{\partial \alpha} - y_{ii} \frac{\partial (q_i^{Norm})^{big}}{\partial \alpha} \ . \tag{52}$$

Отсюда получаем:

$$\frac{\partial y_{ii}}{\partial \alpha} = -\frac{\sum\limits_{j \neq i} y_{ij} \frac{\partial (q_j^{Norm})^{big}}{\partial \alpha} + y_{ii} \frac{\partial (q_i^{Norm})^{big}}{\partial \alpha} + \sum\limits_{j \neq i} \frac{\partial y_{ij}}{\partial \alpha} (q_j^{Norm})^{big}}{(q_i^{Norm})^{big}}. \qquad (53)$$

где производные зарядов находятся из (51).

## 5. Выводы.

В данной работе получены условия нормализации для столбцов (строк) матриц модели COSMO. Из них находятся точные выражения для диагональных элементов основной матрицы COSMO. Для этих целей необходимо решить электростатическую задачу распределения заряда по металлической поверхности. Разработан метод укрупненных элементов для COSMO, аналогичный подобному методу в PCM [1]. Указанный метод позволит приближенно решать уравнение COSMO и находит радиусы Борна для модели SGB[2].

**Приложение.**

### 1. Алгоритм подсчета нормалей по координатам поверхностных элементов.

В моделях COSMO часто не задаются нормали, а лишь координаты поверхностных элементов. В этом случае нормали могут быть найдены на основе следующего метода. Нормаль элемента ищем как векторную среднюю нормалей к треугольным элементам с вершиной в центральной точке многоугольного элемента с весами, пропорциональными площадям треугольников

$$a_{gi} = r_{gi} - r_g, \qquad (54)$$

где $r_g$ – радиус-вектор центра текущего поверхностного элемента, $r_{gi}$ – радиус-вектора центров соседних поверхностных элементов, $i=1,...,N_{eg}$ – номера соседних поверхностных элементов.

Введем следующее обозначение:

$$a_{gi\_next} = \begin{cases} a_{g(i+1)} & \text{for } 1 \le i < N_{eg} \\ a_{g1} & \text{for } i = N_{eg} \end{cases}. \qquad (55)$$

На первом шаге алгоритма.

$$n_g = \frac{\sum_{i=1}^{N_{eg}} ([a_{gi} \times a_{gi\_next}])}{\left| \sum_{i=1}^{N_{eg}} ([a_{gi} \times a_{gi\_next}]) \right|}$$

Нормаль текущего поверхностного элемента:

$$n_g = \frac{\sum_{i=1}^{N_{eg}} ([a_{gi} \times a_{gi\_next}] sign(n_{near}[a_{gi} \times a_{gi\_next}]))}{\left| \sum_{i=1}^{N_{eg}} ([a_{gi} \times a_{gi\_next}] sign(n_{near}[a_{gi} \times a_{gi\_next}])) \right|} \qquad (56)$$

где $n_{near}$ - нормаль соседнего поверхностного элемента, используется для согласования направлений нормалей.

При таком вычислении все нормали будут направлены в одну сторону – либо вовнутрь, либо наружу. Однако нам необходимо, чтобы они все были направлены наружу. Для этого объема поверхности вычисляется по следующей формуле:

$$V = \frac{1}{3}\oint_S (\boldsymbol{r}\cdot\boldsymbol{n})dS \tag{57}$$

где **r** – радиус-вектор текущей точки поверхности, **n** - ее нормаль. Если V<0, то для всех нормалей меняем знак:

$$\boldsymbol{n}_g = -\boldsymbol{n}_g. \tag{58}$$

## 2. Алгоритм нахождения распределения ненулевого заряда $q^{Norm}$ по исходной полости, заполненной металлом, и вакуумом вне нее.

Опишем алгоритм решения «зеркальной» задачи – нахождения распределения заряда по исходной полости, заполненной металлом, и вакуумом вне нее. Напомним четыре условия, которым должно удовлетворять такое распределение заряда:

I) Напряженность электрического поля на поверхности полости (с внутренней стороны) равна нулю.

II) Напряженность электрического поля внутри полости (в том числе в точках, где были размещены точечные заряды) равна нулю.

III) Потенциал внутри полости (в том числе в точках, где были размещены точечные заряды) равен константе $C_{val}$. Потенциал на бесконечности принимаем за ноль

IV) Потенциал на поверхности полости равен той же константе $C_{val}$.

Тогда уравнения, определяемые условиями I) и II), запишутся как условия ортогональности вектора зарядов известной системе векторов.

Из условия I) для направлений, *перпендикулярных* нормали поверхностного элемента:

$$\left[\sum_{i \neq j} q_j^{Norm} \frac{(\boldsymbol{r}_i - \boldsymbol{r}_j)}{|\boldsymbol{r}_i - \boldsymbol{r}_j|^3} \times \boldsymbol{n}_i \right] = 0. \tag{59}$$

Условия I) для направлений *вдоль* нормалей поверхностных элементов уравнения для поверхности:

$$\left( \frac{\partial \varphi_{in}(\boldsymbol{r})}{\partial n} \right) = 0. \tag{60}$$

При разбиении на поверхностные элементы из (2) при (60) и при Q=0 в матричной форме уравнение (2) выглядит следующим образом (аналогично получению уравнения (6) из (5)):

$$- Aq^{Norm} + 4\pi q^{Norm} = 0. \tag{61}$$

То же самое можно записать через матричные элементы $a_{ij}$:

$$- \sum_{j \neq i} a_{ij} q_j^{Norm} - a_{ii} q_i^{Norm} + 4\pi q_i^{Norm} = 0 \tag{62}$$

или в развернутом векторном виде:

$$- \boldsymbol{n}_i S_i \sum_{j \neq i} q_j^{Norm} \frac{(\boldsymbol{r}_i - \boldsymbol{r}_j)}{|\boldsymbol{r}_i - \boldsymbol{r}_j|^3} + q_i^{Norm} \left( \sum_{j \neq i} \frac{(\boldsymbol{r}_j - \boldsymbol{r}_i)}{|\boldsymbol{r}_j - \boldsymbol{r}_i|^3} \boldsymbol{n}_j S_j \right) = 0. \tag{63}$$

Для плоских малых элементов это вырождается в следующее уравнение:

$$- \boldsymbol{n}_i S_i \sum_{j \neq i} q_j^{Norm} \frac{(\boldsymbol{r}_i - \boldsymbol{r}_j)}{|\boldsymbol{r}_i - \boldsymbol{r}_j|^3} + q_i^{Norm} 2\pi = 0. \tag{64}$$

Из условия II) для точек расположения центров атомов получаем:

$$\sum_m \frac{q_m^{Norm}}{|\boldsymbol{R}_i - \boldsymbol{r}_m|^3} (\boldsymbol{R}_i - \boldsymbol{r}_m) = 0 \quad \forall i. \tag{65}$$

Используем итерационный метод для нахождения $q^{Norm}$ из уравнений (58), (62) и (64). Это, по сути, нахождение вектора, ортогонального данному набору векторов.

(1) В качестве начального приближения мы выбираем вектор $q^{(0)Norm} = S$, где $S$ - столбец из площадей поверхностных элементов, и нормируем его на единицу: $\|q^{(0)Norm}\| = 1$

(2) Из условия эквипотенциальности поверхности все заряды поверхностных элементов имеют один знак, который выберем положительным $q^{Norm} > 0$

Отсюда $(q^{(0)Norm})^T \cdot q^{Norm} > 0$, значит, у начального приближения есть ненулевая проекция на искомый вектор, что делает его приемлемым для итераций.

(3) Процесс ортогонализации Грамма – Шмидта [9] будет слишком затратный. Поэтому сделаем процесс поиска итерационным:

а) Нормализуем на 1 все 3N+3M векторов, входящие в (58), (62) и (64). Назовем этот набор векторов $d_k$ $k=1,..., 3N+3M$; $\|d_k\|=1$ D.

Текущий вектор $q_j^{(k-1)Norm}$ будем менять последовательно с каждым вектором из набора:

$q_j^{(k)Norm} = q_j^{(k-1)Norm} - ((q_j^{(k-1)Norm})^T d_k) q_j^{(k-1)Norm}$, $k = 1,...,3N+3M$. Если какая-то из компонент этого вектора получается меньше нуля $(q_j^{(k)Norm})_m < 0$, то обнуляем ее $(q_j^{(k)Norm})_m = 0$.

б) Если $\|q_j^{(3N+3M)Norm}\| \approx 1$, то прекращаем итерации, если нет – возвращаемся на пункт (а) алгоритма с $q_{j+1}^{(0)Norm} = q_j^{(3N+3M)Norm} / \|q_j^{(3N+3M)Norm}\|$.

# Список литературы